\documentclass[aps,prd,floatfix,showpacs,nofootinbib,twocolumn,preprintnumbers,twoside]{revtex4}
\usepackage{amsmath} 
\usepackage{amssymb}
\usepackage{hyperref}
\usepackage{epsfig}
\usepackage{cancel}
\usepackage{epstopdf}
\usepackage{graphics}
\usepackage{graphicx}
\usepackage{multirow}
\usepackage{color}

\newcommand{\met}{\displaystyle{\not{\text{E}}}_T}

\newcommand{\be}{\begin{equation}}
\newcommand{\ee}{\end{equation}}
\newcommand{\bea}{\begin{eqnarray}}
\newcommand{\eea}{\end{eqnarray}}

\begin{document}

\setlength{\pdfpageheight}{\paperheight}
\setlength{\pdfpagewidth}{\paperwidth}
\preprint{IIT-CAPP-13-06, ANL-HEP-PR-13-38}

\title{Dark matter with $t$-channel mediator: a simple step beyond contact interaction}

\author{Haipeng An$^1$, Lian-Tao Wang$^2$, and Hao Zhang$^{3,4,5}$}

\affiliation{$^1$Perimeter Institute, Waterloo, Ontarrio N2L 2Y5, Canada \\
$^2$Kavli Institute for Cosmological Physics and the Enrico
  Fermi Institute, The University of Chicago,~5640 S. Ellis Ave,
  Chicago, IL 60637 \\
$^3$ Illinois Institute of Technology, Chicago, Illinois 60616-3793, USA \\
 $^4$ High Energy Physics Division, Argonne National Laboratory, Argonne, Illinois 60439, USA\\
  $^5$ Department of Physics, University of California, Santa Barbara, California 93106, USA}

\begin{abstract}
Effective contact operators provide the simplest parameterization of dark matter searches at colliders. However, light mediator can significantly change the sensitivity and search strategies. Considering simple models of mediators is an important next-step for collider searches. In this paper, we consider the case of a $t$-channel mediator. Its presence opens up new contributions to the monojet$+\met$ searches and can change the reach significantly. We also study the complementarity between searches for processes of monojet$+\met$ and direct pair production of the mediators. Mediator pair production also gives important contribution to a CMS-like monojet$+\met$ search where a second hard jet is allowed.There is a large region of  parameter space in which the monojet$+\met$ search provides the stronger limit. 
Assuming the relic abundance of the dark matter is thermally produced within the framework of this model, we find that in the Dirac fermion dark matter case, there is no region in the parameter space that satisfies the combined constraint of monojet$+\met$ search and direct detection; whereas in the Majorana fermion dark matter case, the mass of dark matter must be larger than about 100 GeV. If the relic abundance requirement is not assumed, 
the discovery of the $t$-channel mediator predicts additional new physics. 
\end{abstract}

\date{\today}

\maketitle

\section{Introduction}

The identity of dark matter (DM) is one of the central questions in particle physics and cosmology. Many experimental efforts are underway to search for the answer. It is also one of the main physics opportunities of the Large Hadron Collider (LHC).  In recent years, there have been significant progress in using simple effective field theory to combine the results of the LHC searches with limits from direct detection experiments \cite{Cao:2009uw,Beltran:2010ww,Goodman:2010yf,Bai:2010hh,Goodman:2010ku,Goodman:2010qn,Fortin:2011hv,Wang:2011sx,Chen:2011zzu,Rajaraman:2011wf,Shoemaker:2011vi,Fox:2011pm,Fox:2011fx,Haisch:2012kf,Huang:2012hs,Zhou:2013fla,Lin:2013sca}. There have also been earlier studies for similar search channels \cite{Birkedal:2004xn,Petriello:2008pu,Gershtein:2008bf}.

The contact operator approach is based on the simplified assumption that the particles conducting the interaction between DM and the SM particles are heavy, and therefore can be integrated out. The constraints on the energy scale of these effective operators from the LHC searches are around several hundred GeV scale. However, with the ability to probe up to TeV energy scale, the unitarity constraints might be violated at the LHC. As a result, the constraints from contact operator studies cannot be applied directly to UV complete models. 
Therefore, it is useful to  consider the case in which the mediator is lighter and within its energy reach. This would inevitably introduce more model dependence. Therefore, it is useful to consider the simplest extensions first. 

\begin{figure}[!htb]
\includegraphics[scale=0.3]{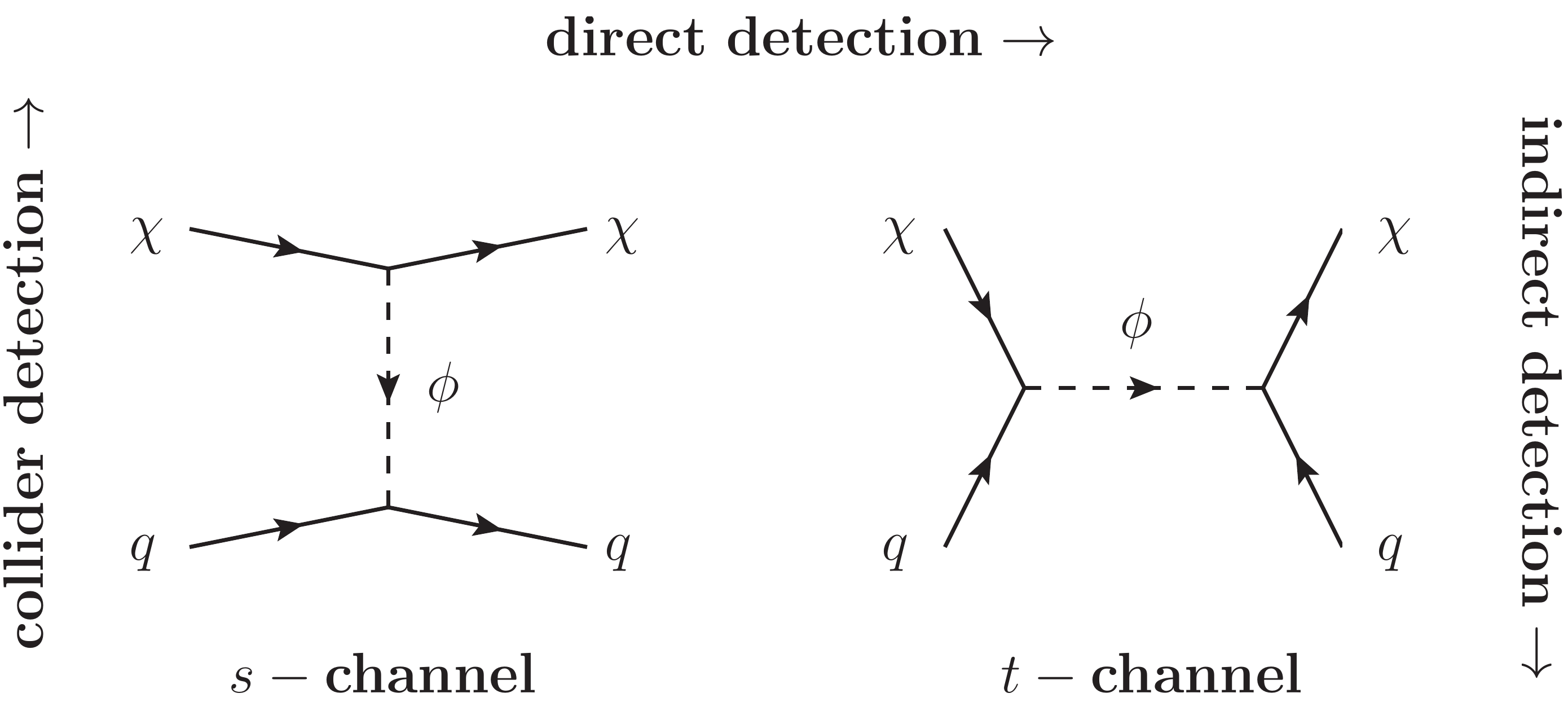}
\caption{\label{fig:SvsT} Diagrams for direct detection mediated by s-channel (left panel) and $t$-channel (mediators).  }
\end{figure}

One such simple scenario is the so-called ``$s$-channel" model, in which the scattering of the DM with nucleus is mediated by the exchange of a mediator particle $\phi$, as shown in the left panel of Fig.~\ref{fig:SvsT}. 
At colliders, it can be produced as a $s$-channel resonance through the $q \bar q \to \phi \to \chi\bar\chi$ process.
Hence, the limit from monojet$+\met$ type searches can be affected significantly. At the same time, direct searches for the resonance $\phi$, such as in the di-jet channel, provide complementary information. This has been demonstrated in the case that the mediator $\phi$ is a massive spin-1 particle \cite{An:2012va,An:2012ue,Frandsen:2012rk}. 

\begin{figure}[!htb]
\includegraphics[scale=0.3]{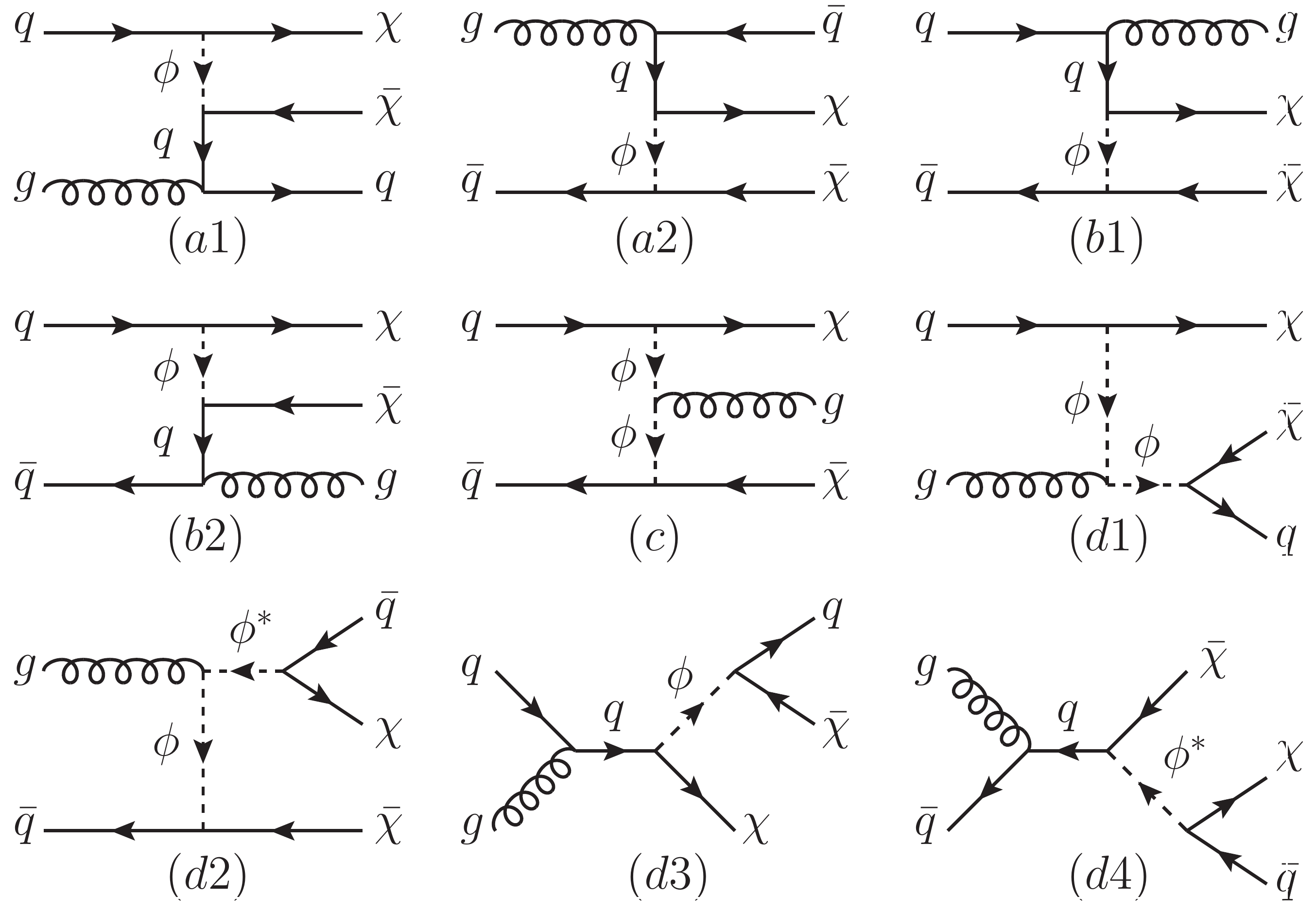}
\caption{\label{fig:monojet} Diagrams for processes of dark matter pair production associated with a single quark or gluon at the LHC in the $t$-channel mediator scenario. (a1,a2) Initial state gluon-split processes; (b1,b2) initial state gluon-emission processes;
(c) gluon-emission from the $t-$channel mediator; (d1-d4) mediator direct production processes. }
\end{figure}

\begin{figure}[!htb]
\includegraphics[scale=0.35]{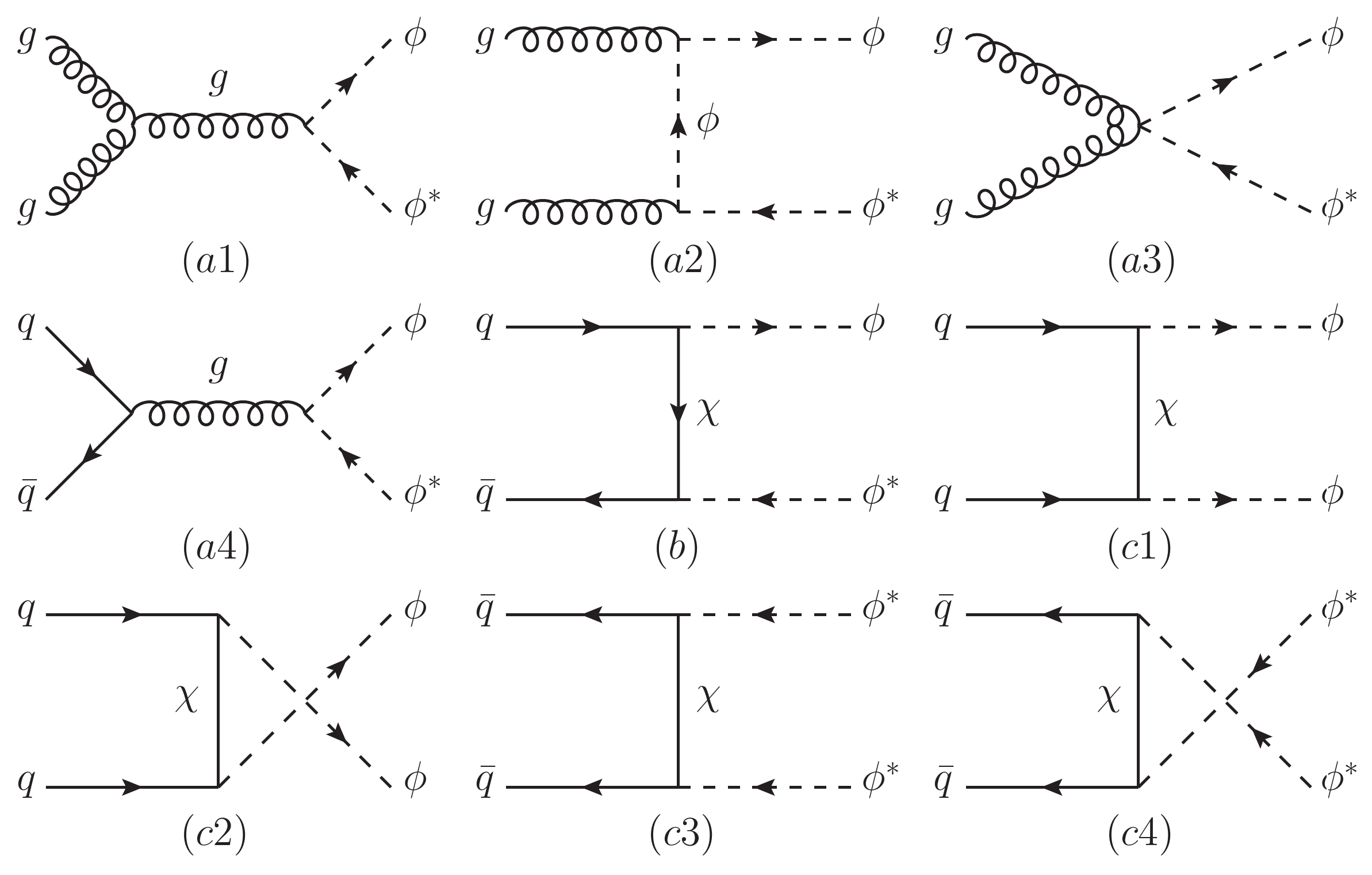}
\caption{\label{fig:di-jet} Diagrams for mediator pair production  processes at the LHC, which leads to di-jet $+\met$ signal. (a1-a4) Diagrams from purely QCD interaction; (b) Diagram from the $t$-
channel DM exchanging; (c1-c4) Diagrams from the $t$-channel Majorana DM
exchanging. }
\end{figure}

In this paper, we consider the other simple possibility in which the DM-nucleus interaction is mediated by going through an intermediate state. We call this the $t$-channel mediator. We focus on the cases that the DM is either a Dirac or Majorana fermion.  In this case, the light mediator also plays an important (and different) role in the collider searches. In particular, it contributes to the monojet$+\met$ searches by being directly produced and decaying into $q+\chi$, as shown in (d1-d4) of Fig.~\ref{fig:monojet}. Moreover, in the most recent monojet$+\met$ search by the CMS collaboration~\cite{CMS-PAS-EXO-12-048} , a second hard jet is also allowed to increase the signal rate. As a result, this search is also sensitive to the di-jet$+\met$ processes, especially in the region where the mediator can be pair-produced. 
At the meanwhile, the process of the pair-production of the mediator is also constrained by squark searches, in which more than two hard jets are triggered~\cite{CMS-PAS-SUS-13-012}. As we will show in this paper, these two channels are complementary. 

This paper is organized as follows. In Section~\ref{sec:framework}, we describe the scenario studied in this paper. In Section~\ref{sec:directdetection}, we discuss the leading direct detection channels. 
In Section~\ref{sec:lhc}, we present the constraints from LHC reaches. In Section~\ref{sec:combine}, 
we study the impact of the assumption that the relic abundance of the DM is thermally produced within the framework of this simple model. In Section~\ref{sec:14TeV}, we present the perspective $5\sigma$ sensitivity of the LHC with 14 TeV center-of-mass energy. 
Section~\ref{sec:summary} contains our conclusion.

\section{Framework}
\label{sec:framework}

In the $t$-channel mediator scenario, we consider interactions of the form
\bea
{\cal L}_{\chi} = \lambda_{q} \bar \chi \phi^* q + h.c. \ ,
\label{eq:lagrangian}
\eea
where $q$, $\chi$ and $\phi$ are the quark field, DM field and the mediator, respectively. For fermionic (scalar) DM, the mediator $\phi$ would be a scalar (fermion). The mediator $\phi$ is also necessarily colored. 

In general, Eq.~(\ref{eq:lagrangian}) may induce flavor changing neutral current which are strongly constrained by flavor experiments. However, these constraints can be avoided by imposing the minimal flavor violation (MFV) structure to the Yukawa couplings~\cite{D'Ambrosio:2002ex}. In the quark sector, without turning on the Yukawa couplings, the SM Lagrangian contains a $U(3)_Q\times U(3)_u\times U(3)_d$ flavor symmetry. Now, for simplicity, let's first assume that $\chi$ is a singlet of the flavor group. Then, to make ${\cal L}_{\chi}$ invariant, the simplest choice is to make $\phi$ to be the $\bf{3}$-representation of one of the three $U(3)$ flavor groups. Therefore, in general, Eq.~(\ref{eq:lagrangian}) can be written as
\begin{eqnarray}
{\cal L}_{\chi} &=& \lambda_{Q} \bar\chi \mathbb{P}_{L}Q\phi_{Q}^{*} + \lambda_{u} \bar\chi \mathbb{P}_{R}u\phi_{u}^{*} + \lambda_{d} \bar\chi \mathbb{P}_{R}d\phi_{d}^{*}  \nonumber \\
&& +  \frac{\lambda_{Qu}^{(1)} \bar\chi H \phi_Q^* Y_u \mathbb{P}_{R} u}{\Lambda} + \frac{\lambda_{Qd}^{(1)} \bar\chi \tilde H \phi_Q^* Y_d \mathbb{P}_{R} d}{\Lambda} \nonumber \\
&& + \frac{\lambda_{Qu}^{(2)} \bar Q H  Y_u \phi_u \mathbb{P}_{R} \chi}{\Lambda} + \frac{\lambda_{Qd}^{(2)} \bar Q \tilde H  Y_d \phi_d \mathbb{P}_{R} \chi}{\Lambda} \nonumber \\
&&+ {\rm h.c.}\ , 
\end{eqnarray}
where $H$ is the Higgs field and $\tilde H = i\sigma_2 H^*$, $Y_u$ and $Y_d$ are the two Yukawa couplings. For the monojet$+\met$ processes, the parton level processes are shown in Fig.~\ref{fig:monojet}, where we can see that the at least one quark or anti-quark initial state is needed. Therefore, all the terms proportional to $Y_u$ or $Y_d$ are in general suppressed by the small masses of the quarks in first two generations. Therefore, in the case that $\chi$ is a $SU(2)$ singlet, to study the generic feature of monojet$+\met$ constraint on the ``$t$-channel" completion of DM models, we can neglect the terms proportional to the Yukawa couplings. Furthermore, the signatures in collider or direct detection experiments are not sensitive to the chirality of the quarks unless $\lambda_{Q,u,d}$ are tuned to have some special relations. Therefore, in this work, in the case that $\chi$ is a SM singlet, we keep only the $\lambda_u$ and $\lambda_d$ terms and assume $\lambda_u = \lambda_d\equiv\lambda$. To simplify our presentation, we also assume that the $\phi_u$ and $\phi_d$ are degenerate that $M_{\phi_u} = M_{\phi_d} \equiv M_\phi$. Then, the Lagrangian can be simplified as 
\begin{equation}\label{lagrangian}
{\cal L}_\chi = \lambda \bar\chi_L q_R \phi^* + {\rm h.c.} \ .
\end{equation} 
For simplicity, we focus on the case that only the right-handed quarks are coupled. For the coupling with left the handed  quarks, minimally, either the mediator or the DM needs to be in a $SU(2)_L$ doublet.   There could be additional signals if the DM is part of a larger multiplet. However, we limit ourselves to the simplest case of singlet DM in this paper.

We assume there are multiple mediators, and they form a multiplet which  has the same flavor content as all the right-handed quarks. Moreover,  all the members of the mediator multiplet are degenerate in mass. A familiar example of this type is the right-handed squarks with universal masses. The possibility of ``flavored" DM has been discussed in Ref.~\cite{Agrawal:2011ze}. In this case, depending the flavor representation of the DM multiplet, it couples to a subset of the left or right-handed quarks. Except for the case in which the DM only couples to top \cite{Kumar:2013hfa}, this case is simply related to the case we study. Of course, as discussed in Ref.~\cite{Agrawal:2011ze}, there are additional signatures in this scenario. Since we focus on the generic features which are common to large class of models, we will not discuss these signals further here. The constraints to this specific $t-$channel model can be 
found in \cite{Garny:2012eb,Garny:2013ama}.

\section{Direct Detection}
\label{sec:directdetection}

In DM direct detection experiments, due to the $\sim$keV scale energy transfer, one can use an effective theory approach to calculate the direct detection signals. 
Integrating out the heavy mediators, at leading order, the effective operator can be written as
\begin{eqnarray}\label{o1}
\mathcal{O}_1 &=& \frac{\lambda^2}{M_\phi^2}\bar\chi_L q_R\bar q_R\chi_L \nonumber \\
                          &=& \frac{\lambda^2}{2 M_\phi^2} \bar\chi_L\gamma_\mu \chi_L \bar q_R \gamma^\mu q_R \ ,
\label{eq:contact_operator}
\end{eqnarray}
where the Fierz transformation has been used in the last step. 
In the case that $\chi$ is a Dirac fermion, the direct detection signal is dominated by the spin-independent (SI) interactions between $\chi$ and nucleus, and the $\chi$-nucleon scattering cross section can be written as
\begin{equation}\label{o1si}
\sigma_{\rm SI}^{\rm (D1)} = \frac{9\lambda^4 \mu^2_{\chi N}}{64\pi (M_\phi^2 - M_\chi^2)^2} \ ,
\end{equation}
where $\mu_{\chi N} = M_\chi M_N/(M_\chi + M_N)$ is the reduced mass of $\chi$ and the nucleon. Spin-dependent (SD) signals can also be induced by ${\cal O}_1$, and the cross section can be written as
\begin{equation}\label{o1sdp}
\sigma_{\rm SD}^{\rm (D1)} = \frac{3\lambda^4 \mu_{\chi N}^2 (\Delta_u^p+\Delta_d^p+\Delta_s^p)^2}{64\pi (M_\phi^2 - M_\chi^2)^2} \ ,
\end{equation}
where $\Delta_q^p$ are defined as $2s_\mu\Delta_q^p = \langle p|\bar q \gamma_\mu \gamma_5 q|p\rangle$ in which $s_\mu$ is the proton spin operator. The values of $\Delta_u^p$, $\Delta_d^p$ and $\Delta_s^p$ can be found in Ref.~\cite{Belanger:2008sj}.   
However, due to the coherent scattering, the SI signal is enhanced by $A^2$ where $A$ is the atomic number of the nucleus.

In the case that $\chi$ is a Majorana fermion, the leading direct detection signal from ${\cal O}_1$ is SD, and the $\chi$-nucleon scattering cross section can be written as
\begin{equation}\label{o1SD}
\sigma_{\rm SD}^{(\rm M1)} = 4 \sigma_{\rm SD}^{(\rm D1)} \ .
\end{equation}
Suppressed SI signals in this case can be generated. Integrating out $\phi$, dimension-7 operators 
\begin{equation}
{\cal O}_2 = \frac{\alpha_S}{4\pi} G^{a\mu\nu} G^a_{\mu\nu} \chi^2 ~{\rm and \ }~ {\cal O}_3 = m_q \bar q q \chi^2 
\end{equation} 
will appear, which lead to SI signals. It is easy to see that if $\chi$ is massless, there is a chiral symmetry which forbids these operators. Therefore, their Wilson coefficients $C_2$ and $C_3$ must be proportional to $M_\chi$. Hence, in the limit that $M_\phi\gg M_\chi + M_q$, at leading order, we have
\begin{equation}
C_2 \sim \frac{\lambda^2 M_\chi}{M_\phi^4} \ , \; C_3 \sim \frac{\lambda^2 m_t^2 M_\chi}{32\pi^2 M_\phi^2 v_{\rm ew}^2 M_h^2}  \ .
\end{equation}
The matrix element of $\left(\alpha_S/4\pi\right) G^{a\mu\nu} G^a_{\mu\nu}$ in the nucleon is proportional to the nucleon mass and comparable to the matrix element of $m_q\bar q  q$. In the region we are interested in, $M_\phi \sim 1$ TeV, we can see that $C_2$ and $C_3$ are of the same order of magnitude. Therefore, the $\chi$-nucleon cross section can be written as
\begin{equation}\label{cxsi2}
\sigma_{\rm SI}^{(2)} \approx \frac{\lambda^4 \mu_{\chi N}^2 }{\pi M_\phi^4}\times0.1 \times \left(\frac{M_N^2}{M_\phi^2}\right) \times \left(\frac{M_\chi^2}{M_\phi^2}\right) \ .
\end{equation}
In the case that $M_\chi$ is comparable to $M_\phi$, the last factor $M_\chi^2/M_\phi^2$ should be changed to an order one parameter.
The details of the calculation can be found in Ref.~\cite{Drees:1992am}.
From Eq.~(\ref{cxsi2}) one can see that for TeV scale $M_\phi$, compared to the usual SI signal, the contributions from $\mathcal{O}_2$ and $\mathcal{O}_3$ are suppressed by a factor of $10^{-6}\sim10^{-7}$, which is comparable to the usually ignored, velocity suppressed contributions. The leading velocity suppressed SI contributions can be found in operator ${\cal O}_1$. Considering only the vector part of the quark current in Eq.~(\ref{o1}), in the non-relativistic limit it matches to the $\chi$-nucleon interaction 
\begin{equation}
\frac{\lambda^2}{8M_\phi^2} \chi^\dagger \gamma_5 \chi N^\dagger N \ .
\end{equation} 
The matrix element of the factor $\chi^\dagger \gamma_5 \chi$ is proportional to the momentum transfer from DM to the targeted nucleus during the collision, whereas the factor $N^\dagger N$ measures the number of nucleons inside the nucleus. Therefore, this contribution is SI and velocity-dependent. Since the velocity of DM is about $10^{-4}\sim10^{-3}$, this contribution is comparable to the SI contributions from $\mathcal{O}_2$ and $\mathcal{O}_3$, especially in the small $M_\chi$ region where the contributions from $\mathcal{O}_2$ and $\mathcal{O}_3$ are further suppressed by $M_\chi^2/M_\phi^2$. 

However, from a simple power counting one can see that both the SI signals from ${\cal O}_1$ or ${\cal O}_2$ are much smaller than the SD signal if the target contains an unsuppressed amount of non-zero spin isotopes. For example, both XENON100 \cite{Aprile:2012nq} and LUX \cite{Akerib:2013tjd} detectors are using liquid xenon as target which contains $^{129}$Xe (spin-1/2) and $^{131}$Xe (spin-3/2) with an an abundance of about 26\% and 21\%, respectively. As a result, if this model does describe the nature of the interaction between DM and the SM particles and DM is a Majorana fermion, we expect the detectors have sensitivity to SD signals to make the first discovery of it. Therefore, in the following discussions, for the case that $\chi$ is a Majorana spinor, we only show the collider limits on SD signals. 

In the case that $\chi$ is a Dirac fermion, the SD signal will be significant if the detector is made of light elements (i.e. hydrogen). But those detectors are only sensitive to low mass DM, which means $M_\chi \ll M_\phi$. In this case, the collider constraint is not sensitive to if $\chi$ is Majorana or Dirac. Therefore, for the Dirac case, we only show the collider limits on SI signals, and the limits for SD signals in the small $M_\chi$ region can be obtained from the limits in the Majorana case using Eq.~(\ref{o1SD}).

\section{LHC searches}
\label{sec:lhc}

Being different from the $s$-channel mediator, the $t$-channel mediators couple to
quarks and color-singlet DM candidate. They can be singly produced associated with a  dark matter particle, leading to a qualitatively new contribution to the mono-jet processes.
For light (lighter than $\sim$ 1 TeV) $t$-channel mediators, the mediators can be pair-produced at the LHC through 
both QCD processes and the DM exchanging processes.  
These processes contribute to signals which are  covered in the squark searches. Moreover, due to the inclusion of a second 
hard jet in the CMS monojet$+\met$ analysis, mediator pair production also gives important contribution to such monojet$+\met$ searches. 

\subsection{Constraints from monojet+$\met$ search}

For monojet$+\met$ searches, the current most stringent constraint is from the search at 8 TeV LHC with 
 a luminosity of 19.5 $\text{fb}^{-1}$ from CMS collaboration \cite{CMS-PAS-EXO-12-048}
\footnote{ATLAS collaboration also publish their result in this channel with 8 TeV $pp$ 
collision, with a lower luminosity of 10$\text{fb}^{-1}$~\cite{ATLAS-CONF-2012-147}}. 
To use their limit, we generate parton level events of $p p\to \chi \chi+{\text{n}}j$ for ${\text{n}}=1,2$ 
using MadGraph5/MadEvent~\cite{Alwall:2011uj}. We use 
CTEQ6L1 parton distribution function (PDF)~\cite{Pumplin:2002vw} 
with 5 flavor quarks in initial states. The parton 
level events are showered using PYTHIA6.4~\cite{Sjostrand:2006za}
and the detector simulation is done by 
PGS4 with anti-$k_T$ jet algorithm with a distance parameter of 0.5. 
The MLM matching scheme is used to avoid double-counting.
We require the signal events to pass the following cuts:
\begin{itemize}
\item At least one central jet which satisfies $p_T>110$ GeV, $\left|\eta\right| <2.4$.
\item At most two jets which satisfy $p_T>30$ GeV, $\left|\eta\right| <4.5$.
\item No isolated electron with $p_T>10$ GeV, $\left|\eta\right| <1.44$ or $1.56<\left|\eta\right| <2.5$.
\item No isolated muon with $p_T>10$ GeV, $\left|\eta\right| <2.1$.
\item $\met>120$ GeV.
\item For events with a second jet, $\Delta\phi_{j_1j_2}<2.5$.
\end{itemize}
Events which pass these cuts are separated in seven signal regions with $\met > 200, 300, 350, 400, 450, 500$, and $550$ GeV. The observed 
upper limit is 4695, 2035, 882, 434, 157, 135 and 131 events for each region
\cite{CMS-PAS-EXO-12-048}. 
In this work, we check all of those seven signal regions. The most stringent constraint is almost always 
from the $\met > 450$ GeV channel. 

The leading order parton level Feynman diagrams with one hard quark or gluon in the final state are shown in Fig.~\ref{fig:monojet}. For the $q\bar q\rightarrow g \chi\chi(\bar\chi)$ process, a gluon can be emitted from both the initial quarks as well as the intermediate $\phi$. In the small $M_\phi$ region, the $qg\rightarrow q\chi\chi(\bar\chi)$ process shown in Fig.~\ref{fig:monojet}(d1-d4) becomes a two-body process. Apart from the enhancement from the phase space, this process benefits from larger parton distribution function of the gluon as well, compared to the anti-quarks in the $q\bar q\rightarrow g \chi\chi(\bar\chi)$ process.
 Therefore, the $qg\rightarrow q\chi\chi(\bar\chi)$ process dominates as long as $\phi$ is relatively light. However, in the larger $M_\phi$ region, the scattering processes from (c) and (d1,d2) are suppressed by $M_\phi^{-2}$, and therefore subdominant.  
At the same time, 
diagrams (d3) and (d4) give the dominant 
contribution, especially when a large jet $p_T$ cut is added.
This is because that the jet from the initial state radiation tends to be soft. 

\begin{figure}[h!]
\includegraphics[scale=0.4]{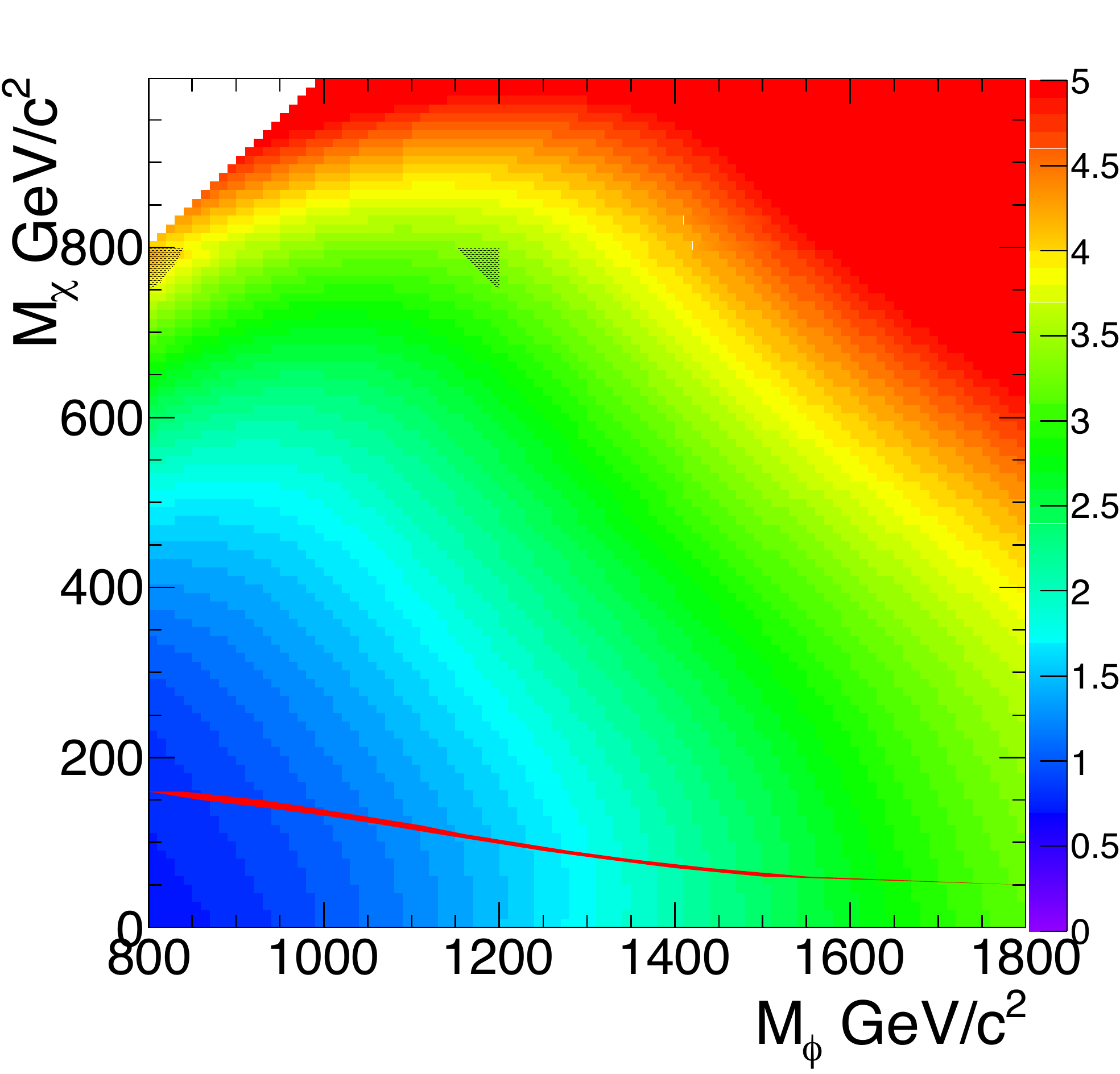}
\includegraphics[scale=0.4]{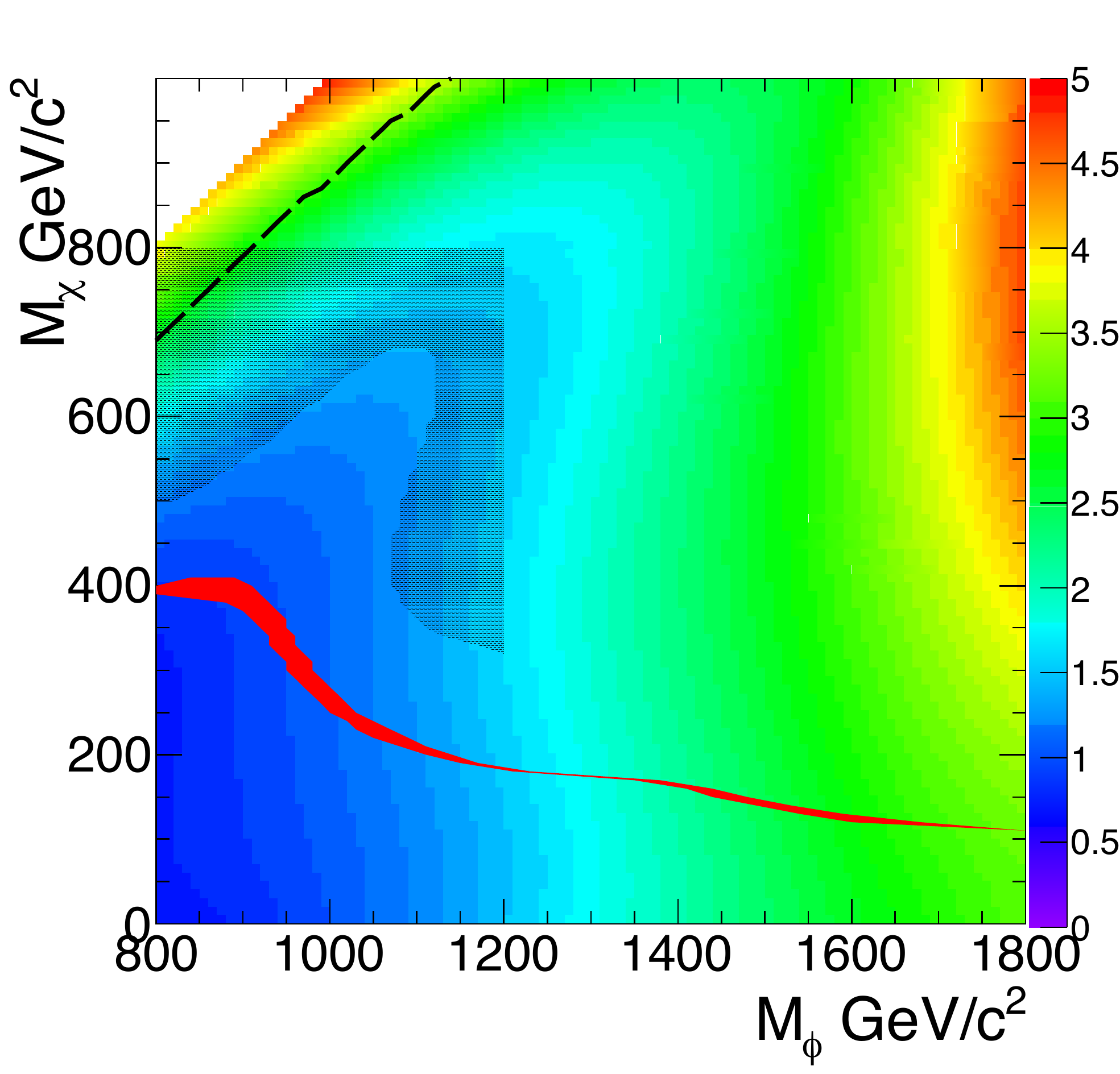}
\caption{The constraints on the $t$-channel mediator model for both the Dirac (upper panel) and 
Majorana (lower panel) cases from the CMS monojet+$\met$ search. The contours are upper limits on the dark matter-mediator-quark coupling $\lambda$. In the lower panel, the region 
above the black dashed curve is excluded by the SD direct detection experiment of the Majorana 
fermion DM. Nearly all of the parameter space of the Dirac fermion DM case is ruled out by the 
direct detection experiments except for very light DM ( $\lesssim6$ GeV ). The red band shows the 
region where the relic abundance of DM can be produced within 3$\sigma$ region of the observed 
value \cite{Ade:2013zuv}. In the shadowed region, the constraint from squark search is stronger than from the monojet+$\met$ 
search (see Fig.~\ref{fig:8tev_squark_search}). 
\label{fig:8tev_mediator}}
\end{figure}

\begin{figure}[h!]
\includegraphics[scale=0.4]{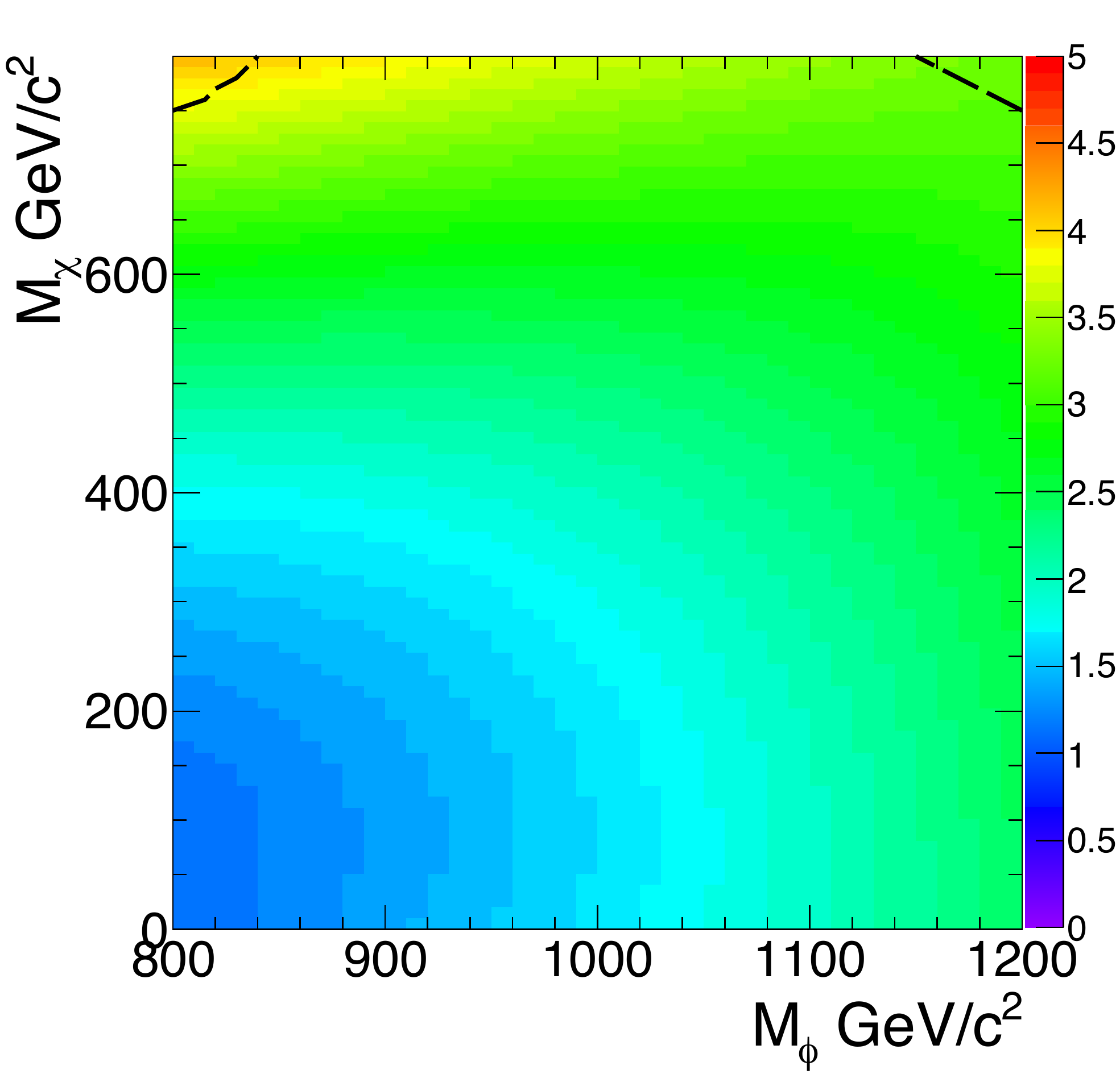}
\includegraphics[scale=0.4]{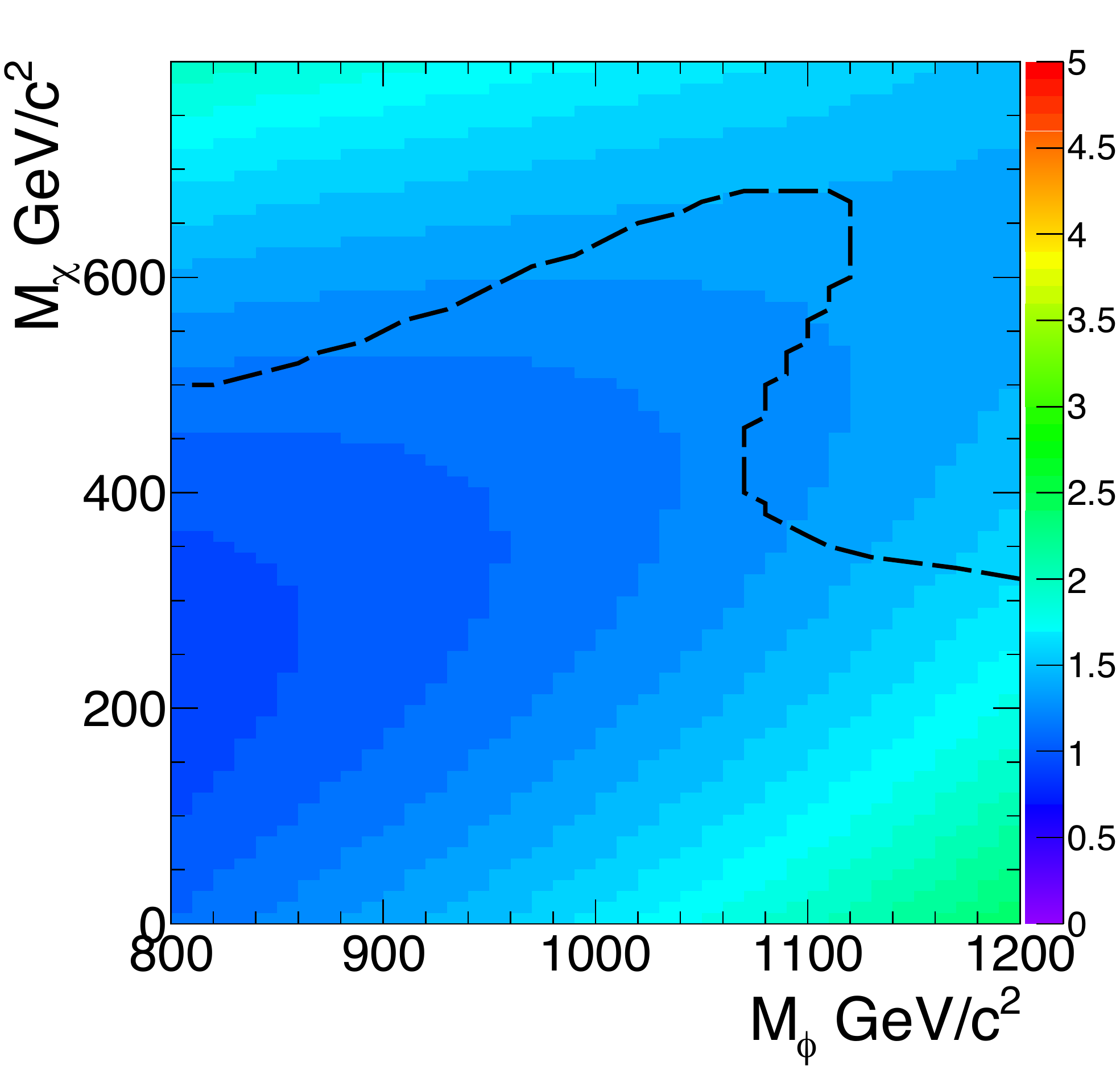}
\caption{The constraints on the $t$-channel mediator model for the Dirac (upper panel)
and Majorana (lower panel) cases from the CMS squark search
at the 8 TeV LHC with 19.5 ${\text{fb}}^{-1}$ integral luminosity. The contours are upper limits on the dark matter-mediator-quark coupling $\lambda$. This constraint is stronger than 
the monojet+$\met$ constraint in the region above the black dashed line.
\label{fig:8tev_squark_search}}
\end{figure}

The leading order parton level diagrams for two hard jets in the final states are the ones with the mediator pair-produced, which are shown in Fig.~\ref{fig:di-jet}. These processes contribute to both the squark and the CMS-like monojet+$\met$ searches. We will focus here on the monojet+$\met$  signal, and discuss the limit from squark searches in the next subsection. 
There are two important contributions to the mediator pair production processes. One is through the QCD production, and the other is through the exchanging of DM particle. In the region where the upper limit of the coupling $\lambda$ is smaller than the coupling of the strong interaction coupling, the QCD process dominates, whereas in the region the constraint on $\lambda$ is weak, the diagrams with exchanging a DM particle dominates. 

The upper limits on the coupling from monojet$+\met$ search for both the Dirac DM and Majorana DM cases are shown in Fig.~\ref{fig:8tev_mediator}. In the Dirac case, for a fixed $M_\phi$, the upper limit on $\lambda$ becomes weaker for larger $M_\chi$. For larger $M_\chi$ not only the phase space becomes smaller, the jet from the decay of $\phi$ to $\chi$ becomes softer as well. From a similar argument, one can see that for a fixed $M_\chi$, as we increase $M_\phi$, the constraint on $\lambda$ becomes stronger at the beginning, then weakens. This effect is more obvious especially in the large $M_\chi$ region.

The Majorana case is qualitatively different from the Dirac case. For fixed $M_\phi$, with the increasing of $M_\chi$, the upper limit on $\lambda$ becomes weaker at the beginning. It becomes stronger in the region where $M_\chi$ is about $M_\phi/2$, and then  weakens again. For example, for $M_\phi \sim 1200$ GeV, there is a strengthening of the limit around $M_\chi \sim 600 $ GeV. This behavior is caused by the exchange of the Majorana $\chi$ in the pair-production process. In the region where $M_\chi$ is relatively large, but not large enough so that the jet from the decay of $\phi$ is too soft, the pair-production process becomes the dominant contribution. Moreover, due to the Majorana property of $\chi$, the contributions from the exchange of $\chi$ is proportional to $M_\chi^2$. Therefore, the production rate becomes larger for larger $M_\chi$.

\subsection{Constraints from squark searches at the LHC}
\label{sec:squark}

The $t$-channel mediators can be copiously produced at the LHC and then decay into a DM particle and a quark. This is very similar to the search in the case of squark search in supersymmetric (SUSY) models. In the case that the gluinos are decoupled.   
The main difference between our scenario and SUSY models is the possibility to enhance the production rate due to the $t$-channel DM exchange process (Fig.~\ref{fig:di-jet}b, c1, c2, c3, c4). Although in the SUSY case, squarks can also be pair-produced through exchanging of neutralinos, the coupling of the squark to neutralino is around the weak coupling. Therefore, this contribution is negligible. However, in the $t$-channel model, we treat the coupling $\lambda$ as a free parameter and it can be quite large. 

Both ATLAS and CMS collaborations show their 95\% C.L. limit to the squark pair
production cross section~\cite{ATLAS-CONF-2013-047,CMS-PAS-SUS-13-012}. We 
calculate the total cross section of $pp\to \phi\phi^* (\phi\phi, \phi^*\phi^*)$ processes and using their
unfolding result to estimate the bound from squark searching at 8 TeV LHC. The 
result from CMS collaboration~\cite{CMS-PAS-SUS-13-012} 
gives a stronger constraint.
The total cross section is calculated using CalcHEP \cite{Belyaev:2012qa}. 
The NLO QCD correction is shown 
to be small for such processes~\cite{Beenakker:2011fu}. 
A typical value of the $K$-factor is smaller than 1.05.
We will neglect it in our calculation. 

The parton-level Feynman diagrams are shown in Fig.~\ref{fig:di-jet}. (a1), (a2), (a3) and (a4) depend only on the strong interaction, whereas (b), (c1), (c2), (c3) and (c4) are mediated by $\chi$ and depend on $\lambda$. The contribution from (c1), (c2), (c3) and (c4) must be proportional to the Majorana mass of $\chi$ since the fermion number is changed and vanishes if $\chi$ is Dirac fermion. 
The constraints from CMS squark search for both the Dirac and Majorana cases are shown in Fig.~\ref{fig:8tev_squark_search}.  In the Majorana case, in the small $M_\chi$ region, the constraint is stronger with larger $M_\chi$, this is because the production rate is proportional to $M_\chi^2$ due to the Majorana nature of $\chi$. The constraint becomes weaker as $M_\chi$ approaches to $M_\phi$ since the jets from the decay of $\phi$ become softer.  Compared to the constraint from monojet$+\met$ search, the constraint from squark search is weaker in most of the parameter region under consideration, especially those with smaller $M_\chi$.

\begin{figure}[h!]
\begin{center}
\includegraphics[scale=0.4]{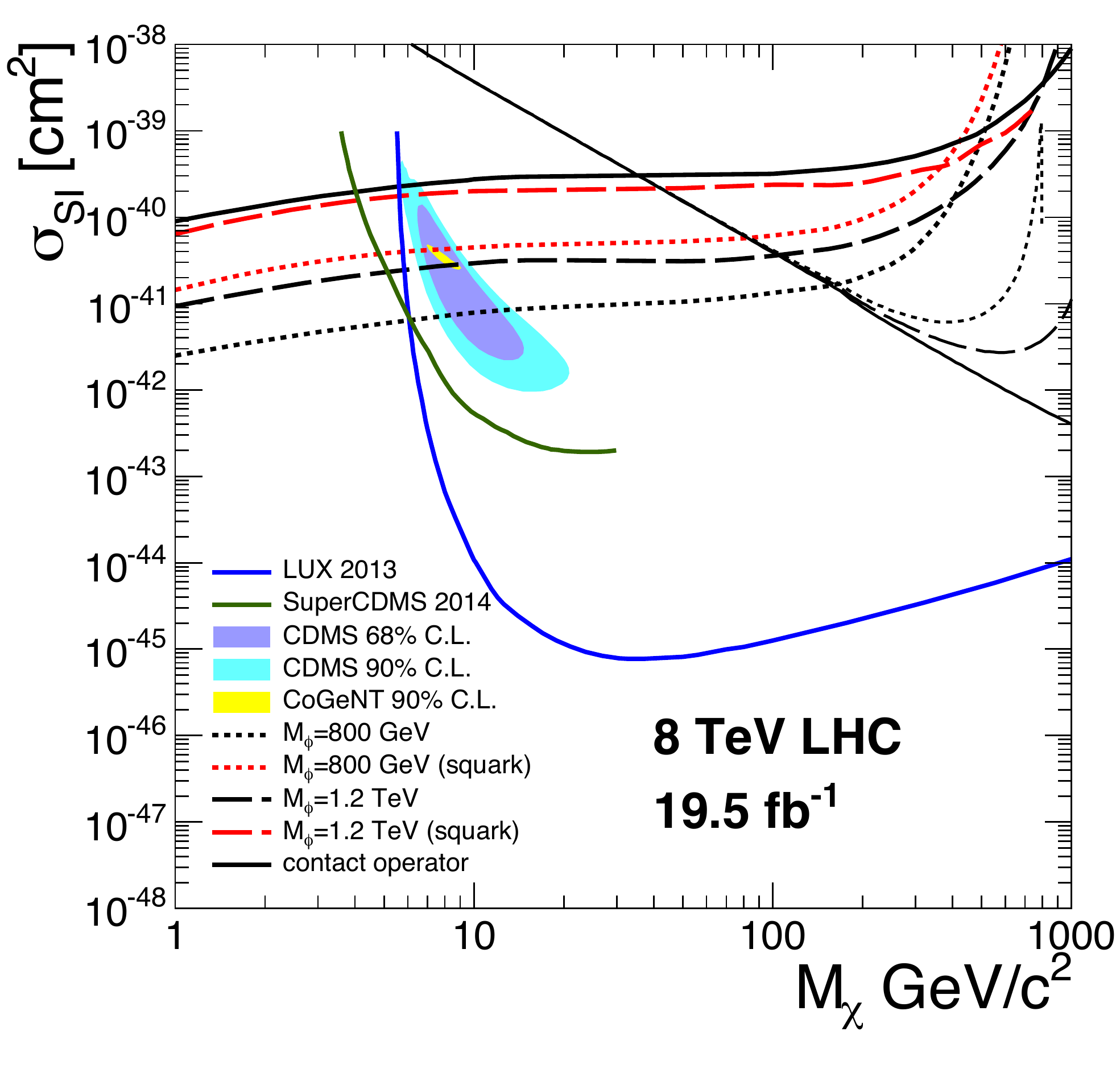}
\includegraphics[scale=0.4]{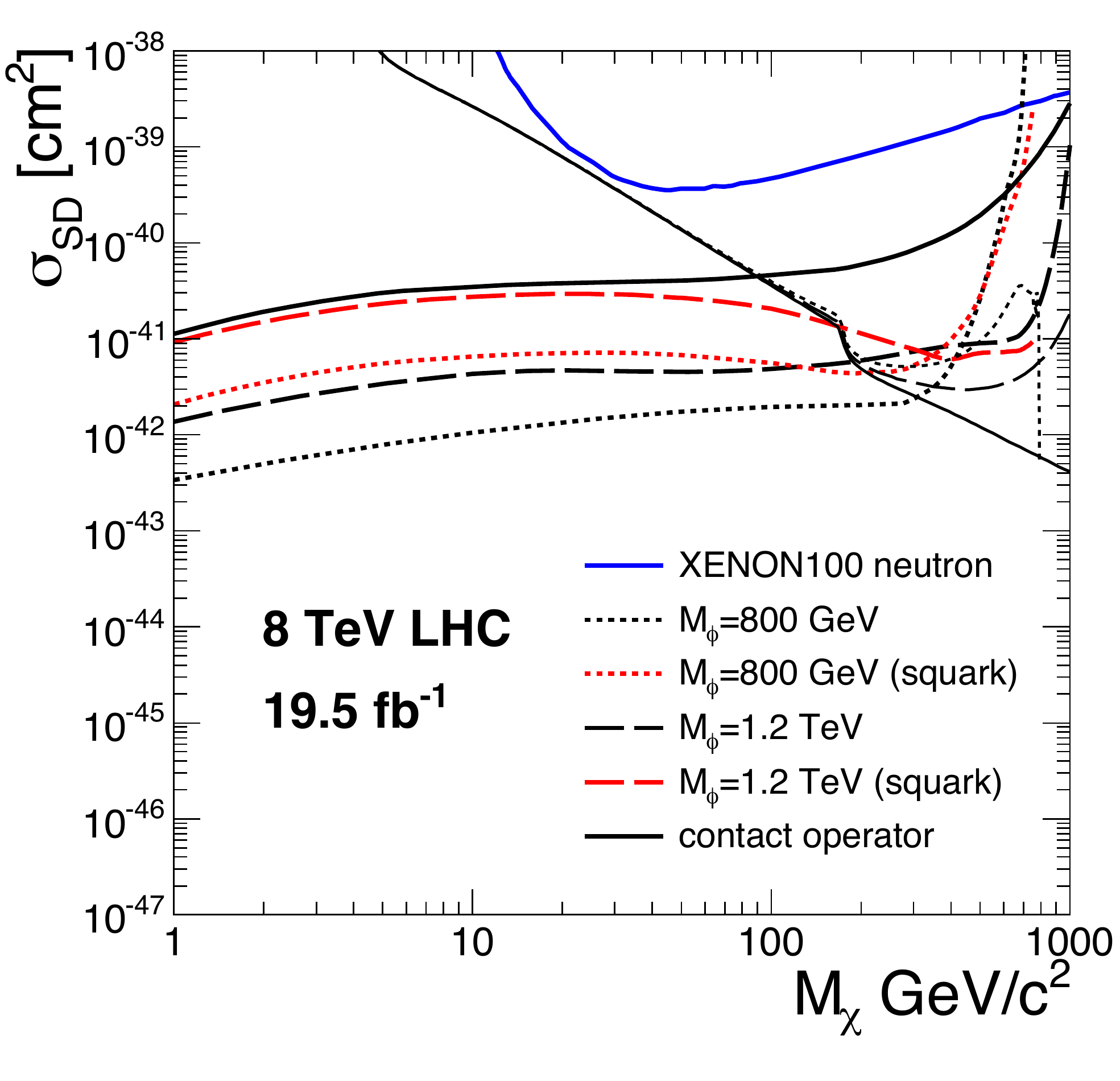}
\caption{ Constraints from monojet+$\met$ and di-jet+$\met$ on direct detection cross sections for both the Dirac (upper) and Majorana (lower) DM cases, for 8 TeV LHC with 19.5 ${\text{fb}}^{-1}$ integral luminosity.
 The constraints from the relic abundance assuming that the model is the unique source for the interaction between DM and SM particles are also shown. For the Dirac DM case, the region relates to the potential WIMP signal from CDMS experiment \cite{Agnese:2013rvf} and CoGeNT experiment \cite{Aalseth:2012if}
 is also shown together with the exclusion region from the first result from LUX
 \cite{Akerib:2013tjd} and SuperCDMS \cite{Agnese:2014aze}. For the Majorana DM case, the constraint from XENON100 \cite{Aprile:2013doa} is shown. 
\label{fig:compare8_14}}
\end{center}
\end{figure}

\section{Combining LHC searches with direct detection and thermal relic abundance}
\label{sec:combine}

Using Eqs.~(\ref{o1si}), (\ref{o1sdp}) and (\ref{o1SD}) the upper limits on $\lambda$ can be translated into upper limits on direct detection cross sections, which are shown in Fig.~\ref{fig:compare8_14}, from which one can see that in the Dirac DM case, the constraint from collider search becomes stronger than the constraint 
from the direct detection experiments only in the region where $M_\chi$ is smaller than about 6 GeV. In the Majorana DM case, however, due to the lack of the enhancement from coherence in the direct detection, the LHC constraint is stronger up to a few hundred GeV. For the monojet+$\met$ constraint, one can see that it becomes much weaker when $M_\chi$ approaches $M_\phi$. This is because in the dominant $qg\rightarrow q\chi\chi(\bar\chi)$ channel, the jet from the decay of $\phi$ becomes soft in this region and needs a large boost to pass the cut. Therefore, in this region, the monojet$+\met$ process is either suppressed by the parton luminosity or by the phase space. Of course, this is the region of the parameter space well covered by the direct detection experiment. On the other hand, This also explains that in the large $M_\chi$ region, the constraint is weaker for smaller $M_\phi$. Therefore, the contact operator approximation underestimates the monojet$+\met$ constraint in the small $M_\chi$ region, but overestimates in the large $M_\chi$ region. In the region that $M_\chi \ll M_\phi$, the collider constraint is not sensitive to $M_\chi$. 
On the other hand, for the constraint from the squark search, in the Majorana case, due to the $M_\chi$ enhancement, the limits can be stronger for large $M_\chi$ region as shown by the red curves in the lower panel of Fig.~\ref{fig:compare8_14}.   

The interesting regions of the recently reported potential light DM signal in CDMS experiment and the anomalies observed by CoGeNT experiment are also shown in Fig.~\ref{fig:compare8_14}. In particular, in the Dirac DM case, the sensitivity of the 8 TeV monojet$+\met$ search is already sensitive to this region. In the Majorana case, since $^{73}$Ge (spin-9/2) only makes up 7.73\% of natural Ge and $^{29}$Si only makes up 4.68 of natural Si. The SD signals from the CDMS and CoGeNT detectors are highly suppressed, and therefore are expected to be deeply inside the exclusion region of the monojet$+\met$ search. 

If we further assume that the relic abundance of the DM are thermally produced within the framework of this simple model (\ref{lagrangian}), 
the thermal annihilation of DM in the early universe is dominated by the quark-anti-quark channels. Assuming the $\chi$ composes all the DM observed in the Universe, the lower limits on direct detection cross sections are shown as the thin black curves in FIG.~\ref{fig:compare8_14}  for Dirac and Majorana DM, respectively. From  FIG.~\ref{fig:compare8_14}, we can see that the limits in the Majorana DM case is more sensitive to the quark mass thresholds. This is because that in the Majorana case, the $s$-wave annihilation cross section is proportional to $m_q^2$, where $m_q$ is the mass of the outgoing quarks. This property can be understood using the effective theory approach. For non-relativistic Majorana DM pairs, we have $\langle0|\bar\chi\gamma_\mu\gamma_5\chi|\chi\chi\rangle \sim k_\mu + {\cal O}(v^1)$, where $k$ and $v$ are the total momentum and relative velocity of the DM pair respectively. In the thermal annihilation case, the DM can only annihilate into quarks with masses smaller than $M_\chi$, so the quark masses can no longer be neglected. The derivation of the right-handed quark current can be written as
\begin{equation}
\partial_\mu \bar q_R\gamma^\mu q_R = m_q \bar q i\gamma_5 q + {\rm anomaly~terms} \ ,
\end{equation}
where the contribution from the anomaly terms leads to the annihilation to the gauge boson final states are loop suppressed and can be neglected in the thermal annihilation process. This contribution can be identified in the process of the annihilation of neutralino into gluons discussed in Ref.~\cite{Drees:1993bh}. In this work, the relic abundance is simulated using micrOmegas3.0~\cite{Belanger:2006is}. 

From Fig.~\ref{fig:compare8_14}, we can see that in the case that $\chi$ is a Dirac fermion, the region allowed by both the LHC searches and the direct detection is not consistent with the requirement of  relic abundance.  Therefore, in this case, this simple model cannot be seen as a complete model in describing the DM interaction with SM particles. There must be other channels for DM to annihilate into SM particles. Of course, the monojet$+\met$ and
squark search channels can still be the leading channel to discover DM at the LHC. 
On the other hand, if $\chi$ is Majorana fermion, Fig.~\ref{fig:compare8_14} shows that if we assume this simple model describes the interaction between DM and SM particles, depending on $M_\phi$, the mass of DM should be larger than around 100 GeV.  Otherwise, there will be additional new physics to look for as well. We also notice that if the relic abundance was generated through this model, the constraint from the monojet search is stronger than from the squark search. 

In the region where $M_\chi$ is close to $M_\phi$, the intermediate $\phi$ approaches its mass shell in this region and therefore enhances the direct detection rate. However, on the 800 GeV curves for both the Dirac and Majorana cases, a sharp turning point appears when $M_\chi$ approaches $M_\phi$. This is because, in this region the co-annihilation channels ({\it e.g.} $\chi \phi \rightarrow q W$) and hidden channels ({\it e.g.} $\phi \phi^*\rightarrow q \bar q,gg$) are open, and the effective annihilation rate gets enhanced. On the 1200 GeV curves, these returning points don't appear since $M_\phi$ is too large and the annihilation rate only through the co-annihilation channels and hidden channels is still not enough to get the correct relic abundance, and a sizable direct annihilation rate is still needed. 

\section{14 TeV LHC perspectives}
\label{sec:14TeV}

\begin{figure}[h]
\begin{center}
\includegraphics[scale=0.4]{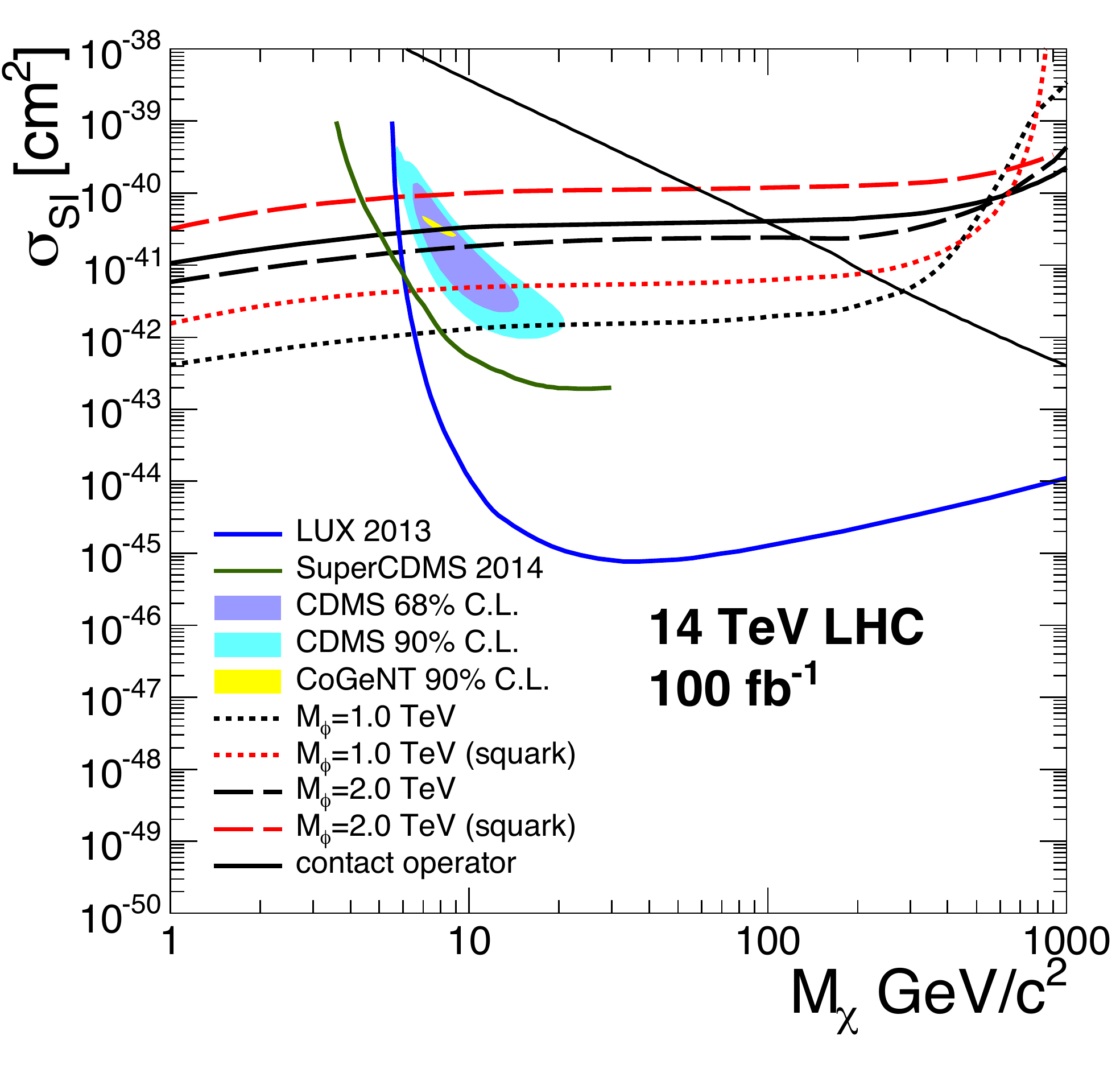}
\includegraphics[scale=0.4]{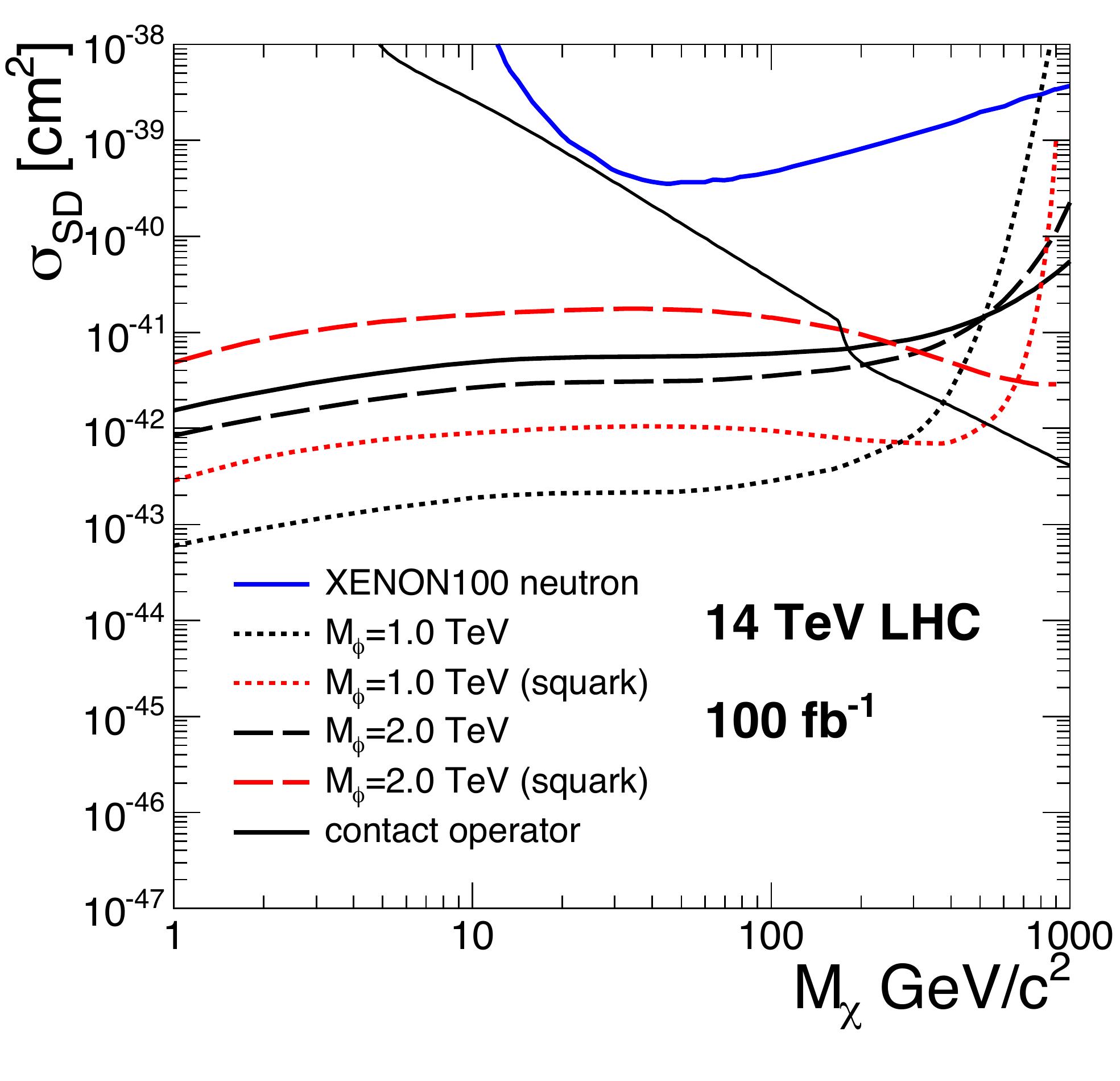}
\caption{ 5$\sigma$ reaches of the monojet+$\met$ and di-jet+$\met$ channels on direct detection cross sections for both the Dirac (upper) and Majorana (lower) DM cases, for 14 TeV LHC with 100 ${\text{fb}}^{-1}$ integral luminosity. The curves for direct detections and relic abundances are the same as in Fig.~\ref{fig:compare8_14} 
\label{fig:compare14}}
\end{center}
\end{figure}

To be complete, we also present the $5 \sigma$ reaches of the monojet$+\met$ channel and the squark search channel at the 14 TeV LHC. For the monojet$+\met$ search, the background is simulated in Ref.~\cite{Vacavant:2001sd} requiring that the $p_T$ of the leading jet and $\met$ larger than 500 GeV. The SM background at the luminosity of 100 fb$^{-1}$ is about $B_{14} \approx 2\times10^4$. For the expected 5$\sigma$ reach, we require that the signal at 100 fb$^{-1}$ larger than $5\sqrt{B}$. The $5\sigma$ reach results for $M_\phi = 1$ and 2 TeV  are shown in Fig.~\ref{fig:compare14}. For $m_\phi\sim 1$ TeV, the 14 TeV LHC can cover most of the interesting region where anomalies from direct detection experiments are reported. 
For a heavy enough 
mediator, both results show good agreement with the contact operator limit. For the squark search, we consider the di-jet$+\met$ channel. We use MadGraph/Event5, PYTHIA6 and PGS to simulate the SM background. For the signal, we use MadGraph/Event5 and PYTHIA6 to generate parton level events and do the parton shower. Then we use FASTJET3.0.0~\cite{Cacciari:2011ma} to do the collider simulation. For the signal region, we require that $\met > 250$ GeV, $p_T(j_1) > 200$ GeV, and $p_T(j_2)>$ 130 GeV, where $j_1$ ($j_2$) is the leading (subleading) jet. The $5\sigma$ results for $M_\phi=1$ and 2 TeV are shown in Fig.~\ref{fig:compare14}, where one can see that the qualitative features of the curves are the same as in the case of the 8 TeV LHC.

\bigskip
\section{Summary and discussions}
\label{sec:summary}

It is likely that the interactions between DM particles and SM particles are mediated by weak  scale physics. Monojet$+{\met}$ process has been proposed to study the properties of the interaction at the LHC. Due to the large energy of LHC, the mediator can be produced directly, and a contact interaction approach may not be a good approximation and violates the unitarity bounds in some cases. Therefore, a UV complete model is needed. In this paper,  we study a simplified $t$-channel UV completion model where the interaction between DM and SM particles are mediated by colored mediators couples to the DM particle and the right-handed quarks.

In this scenario, the relevant processes at the LHC are dark matter pair production associated with a quark or gluon, mediator-dark matter associated production, mediator pair production. Obviously, the first two will give rise to monojet$+\met$ signal, and the last one will be similar to squark pair production. However, since the CMS monojet$+\met$ search also allow second hard jet, the mediator pair production process also gives important (and sometimes even dominant) contribution to this search. In fact, we observe that, in comparison with the squark searches, the CMS-like monojet$+\met$ search gives stronger constraints in most of the parameter space. \footnote{This has also been noticed recently in \cite{Papucci:2014iwa}. }

If the DM particle is Dirac fermion, the dominant direct detection signal is SI, and the monojet$+\met$ search starts to be sensitive to the interesting parameter space in the small $M_\chi$ region. In almost all of the parameter region under consideration, CMS monojet$+\met$ search gives the stronger constraints than the squark search.   In the case that the DM particle is Majorana fermion, the dominant direct detection signal is SD, and the monojet$+\met$ signal is stronger in the region that $M_\chi$ is smaller than a hundred GeV, and 
the squark search is more significant for heavier DM. 

If we further require that the relic abundance of DM in the Universe is generated within the context of this model, in the Dirac DM case, there is no region in the parameter space that reconciles the combined constraint of monojet$+\met$ search and direct detection with constraint from not over closing the universe; and in the Majorana case, the mass of DM must be larger than about 100 GeV. Of course, in both cases, even if the relic abundance requirement can not be satisfied, the monojet$+\met$ and squark searches can still be the leading channels to discover the DM at the LHC. It would be just an indication that there will be more new particles to look for. 

In the Majorana case, inside our galaxy, the p-wave annihilation channel in suppressed either by the velocity. At the meanwhile, if the DM particle couples only to the light quarks, the two-body annihilation channel is suppressed by the light quark masses. In this case, the three-body Internal bremsstrahlung processes dominate the annihilation, which can potentially be detected in the indirect detection experiments~\cite{Garny:2013ama}.   

We end our conclusion with a brief discussion on the connection to Higgs invisible width. 
In this specific model, the process for Higgs decays into a pair of DM particles can be induced at one-loop. Since the DM is assumed to be a SM singlet, this process is predictable within the context of this simple model. Since the Higgs coupling changes the chirality of the quark, and we assume that $\chi$ couples only to the right-handed quarks, the chirality of the quark in the internal line needs to be changed for two times. Therefore, the effective coupling is proportional to $m_q^2$ and negligible for light quarks. The top quark induced effective coupling can be written as
\begin{equation}
{\cal L} \sim \frac{\lambda^2 m_t^2 M_\chi}{32\pi^2 M_\phi^2 v_{\rm ew}} h \bar\chi\chi \ ,
\end{equation}
where $v_{\rm ew} = 246$ GeV is the Higgs vev. In order for Higgs to decay into a pair of DM, $M_\chi$ must be smaller than $M_h/2$, where $M_h = 126$ GeV is the mass of the Higgs boson. From Fig.~\ref{fig:8tev_mediator} one can see that the $M_\phi/\lambda$ mush be smaller than about 500 GeV, Therefore, in this model the effective coupling can be written as 
\begin{equation}
\frac{\lambda^2 m_t^2 M_\chi}{32\pi^2 M_\phi^2 v_{\rm ew}} \approx 6\times10^{-3}\left( \frac{500{\rm~GeV}}{M_\phi/\lambda} \right)^2 \left( \frac{2M_\chi}{M_h} \right) \left(\frac{m_b}{v_{\rm ew}}\right) \ ,
\end{equation}
which is much smaller than the Higgs coupling to the bottom quark, and therefore is not contained by the limit from invisible Higgs decay derived from current LHC data.

\begin{acknowledgments}We thank Maxim Pospelov and Itay Yavin for useful discussions. H.A.'s research at Perimeter Institute is supported by the Government of Canada through Industry Canada and by the Province of Ontario through the Ministry of Research
and Innovation. L.T.W. is supported by the NSF under
grant PHY-0756966 and the DOE Early Career Award under grant. de-sc0003930. 
H. Z is supported by the U.S. DOE under the Grant No. DE-AC02-06CH11357, 
the Grant No. DE- FG02-94ER40840 and the Grant No. DE-FG02-91ER40618. 
\end{acknowledgments}

{\it Note added.} While the work is being complete, Ref.~\cite{Chang:2013oia} appeared which has overlap with the content of this paper.

\bibliography{draft}

\end{document}